\documentclass[twocolumn,showpacs,preprintnumbers,nofootinbib,prd,superscriptaddress,groupedaddress,10pt]{revtex4-1}

\usepackage{graphicx,amssymb,amsmath,amsthm,amsfonts,epsfig,epsf}
\usepackage[linktocpage]{hyperref}
\usepackage[usenames]{color}
\usepackage{epstopdf}

\usepackage{float}
\usepackage{aas_macros}
\usepackage{bm}
\usepackage{dcolumn}
\usepackage[latin1]{inputenc}
\usepackage{latexsym}
\usepackage{rotating}
\usepackage{hyperref}
\usepackage{color}
\usepackage{longtable}
\usepackage{enumerate}
\usepackage{tensor}
\usepackage{url}
\newcommand{\ben}{\begin{enumerate}}
\newcommand{\een}{\end{enumerate}}

\def\be{\begin{equation}}
\def\ee{\end{equation}}
\def\bea{\begin{eqnarray}}
\def\eea{\end{eqnarray}}

\newcommand{\beq}{\begin{eqnarray}}
\newcommand{\eeq}{\end{eqnarray}} 
\newcommand{\ba}{\begin{align}}
\newcommand{\ea}{\end{align}}

\begin{document}

\title{Two worlds collide: Interacting shells in AdS spacetime and chaos}

\author{
Richard Brito$^{1}$,
Vitor Cardoso$^{1,2,3}$, % \footnote{Electronic address: vitor.cardoso@tecnico.ulisboa.pt},
Jorge V. Rocha$^{4}$ % \footnote{Electronic address: jvrocha@icc.ub.edu},
}
\affiliation{${^1}$ CENTRA, Departamento de F\'{\i}sica, Instituto Superior T\'ecnico -- IST, Universidade de Lisboa -- UL,
Avenida Rovisco Pais 1, 1049 Lisboa, Portugal}
\affiliation{${^2}$ Perimeter Institute for Theoretical Physics, 31 Caroline Street North
Waterloo, Ontario N2L 2Y5, Canada}
\affiliation{${^3}$ Dipartimento di Fisica, ``Sapienza'' Universit\`a di Roma \& Sezione INFN Roma1, P.A. Moro 5, 00185, Roma, Italy}
\affiliation{${^4}$ Departament de F\'isica Fonamental, Institut de Ci\`encies del Cosmos (ICCUB), Universitat de Barcelona, Mart\'i i Franqu\`es 1, E-08028 Barcelona, Spain.}

\begin{abstract}
We study the simplest two-body problem in asymptotically anti-de Sitter spacetime: two, infinitely thin, concentric spherical shells
of matter. We include only gravitational interaction between the two shells, but we show that the dynamics of this system 
is highly nontrivial. We observe prompt collapse to a black hole, delayed collapse and even perpetual oscillatory motion,
depending on the initial location of the shells (or their energy content). The system exhibits critical behavior, 
and we show strong hints that it is also chaotic.
\end{abstract}

%\tableofcontents
%\end{widetext}
%\clearpage

%95.35.+d 	Dark matter (stellar, interstellar, galactic, and cosmological) (see also 95.30.Cq Elementary particle processes; for brown dwarfs, see 97.20.Vs; for galactic halos, see 98.35.Gi or 98.62.Gq; for models of the early %Universe, see 98.80.Cq)
%14.80.-j 	Other particles (including hypothetical)
%11.10.St 	Bound and unstable states; Bethe-Salpeter equations
%12.60.-i 	Models beyond the standard model (for unified field theories, see 12.10.-g)
%04.25.D-    Numerical relativity
%04.25.dc    Numerical studies of critical behavior, singularities, and cosmic censorship
%04.25.dg    Numerical studies of black holes and black-hole binaries
%04.25.-g    general relativity: approximation methods, equations of motion
%04.40.-b 	Self-gravitating systems; continuous media and classical fields in curved spacetime
%04.50.-h    Higher-dimensional gravity and other theories of gravity
%04.50.Cd    KaluzaKlein theories
%04.50.Gh    Higher-dimensional black holes, black strings, and related objects
%04.60.Cf    Gravitational aspects of string theory
%04.70.-s    Physics of black holes
%04.70.Bw    Classical black holes
%04.70.Dy    Quantum aspects of black holes, evaporation, thermodynamics
%04.80.-y    Experimental studies of gravity
%04.80.Cc    Experimental tests of gravitational theories
%11.25.Mj    Compactification and four-dimensional models
%11.10.Kk    Field theories in dimensions other than four

\maketitle

%\tableofcontents

%%%%%%%%%%%%%%%%%%%%%%%%%%%%%
%%%%%%%%%%%%%%%%%%%%%%%%%%%%%%%%%%%%%%%
\section{Introduction\label{sec:Intro}}
%%%%%%%%%%%%%%%%%%%%%%%%%%%%%%%%%%%%%%%

Gravitational physics in anti-de Sitter (AdS) space has staged many efforts over the past two decades, mainly driven by the celebrated AdS/CFT correspondence~\cite{Maldacena:1997re,Witten:1998qj,Gubser:1998bc}. However, the first serious study of AdS physics dates back to 1978, when the quantization of scalar fields in such a spacetime was considered~\cite{Avis:1977yn}.
In the early 1980s, research on AdS gravity was propelled mainly by 
investigations of gauged supergravity, where AdS often arises as a supersymmetric vacuum~\cite{deWit:1982bul};
studies of thermodynamics and phase transitions exhibited by black holes (BHs) in AdS~\cite{Hawking:1982dh};
and the development of a Hamiltonian formalism for asymptotically AdS spacetimes~\cite{Henneaux:1985tv}.

The chief feature of asymptotically AdS spacetimes is their confining nature: fields propagating in AdS feel a potential that diverges asymptotically, and light rays reach infinity in finite time. This is of course related to the fact that AdS is not globally hyperbolic, and therefore, a well-posed initial value problem requires that boundary conditions for fields at infinity must be provided~\cite{Friedrich:1995vb,Ishibashi:2004wx}.
This particularity of AdS is at the heart of several interesting recently uncovered phenomena: 
the so-called turbulent instability of AdS~\cite{Bizon:2011gg,Maliborski:2012gx,Dias:2011ss,Dias:2012tq,Garfinkle:2011tc,Buchel:2012uh,Balasubramanian:2014cja,Buchel:2014xwa,Maliborski:2013jca,Maliborski:2014rma,Bizon:2015pfa,Okawa:2013jba,Okawa:2014nea,Okawa:2015xma,Craps:2014vaa,Craps:2014jwa,Craps:2015iia,Craps:2015xya,Craps:2015jma,Dimitrakopoulos:2015pwa,Freivogel:2015wib},
asymptotically AdS solutions such as boson stars, geons and hairy BHs~\cite{Astefanesei:2003qy,Buchel:2013uba,Basu:2010uz,Dias:2011at,Dias:2011tj,Dias:2011ss,Dias:2012tq}, and 
holographic studies of the equilibration of strongly coupled plasmas~\cite{Danielsson:1999fa,Chesler:2008hg,Bhattacharyya:2009uu,Chesler:2010bi,Garfinkle:2011hm,Heller:2012km,Heller:2012je,Liu:2013qca,vanderSchee:2013pia}
and of quantum revivals~\cite{Abajo-Arrastia:2014fma,daSilva:2014zva}.
Most, if not all, of these investigations rely either on intense numerical work or on cumbersome perturbative calculations.

It has been pointed out very recently~\cite{Cardoso:2016wcr} that a system comprised of multiple spherically symmetric (and concentric) pressurized thin shells in a flat space cavity displays extremely rich---yet easily solvable---dynamics. For example, depending on initial conditions it is possible to obtain perpetually oscillating configurations or delayed collapse into a BH, in addition to prompt collapse. Moreover, this setting also exhibited critical behavior as present in the original studies of gravitational collapse~\cite{Choptuik:1992jv,Gundlach:2002sx}---in particular, sharing striking similarities with the most recent analyses in Refs.~\cite{Olivan:2015fmy,Cai:2016yxd}.
For each shell, the problem simply amounts to integrating the motion of a particle in a one-dimensional potential. The shells were assumed to cross without any further interaction besides their gravitational attraction; in other words, they are ``transparent.''

In fact, the dynamics of two or more spherical thin shells have been studied for more than 30 years, albeit in different contexts~\cite{Dray:1985yt,Nunez:1993kb,Miller:1997}. 
For example, Miller and Youngkins~\cite{Miller:1997} studied the chaotic behavior of two concentric, spherical thin shells enclosed by an inner and an outer barrier in the Newtonian regime. Evidence for chaotic motion was also found in Ref.~\cite{Barkov:2001eg}, for a Newtonian system with two shells surrounding a central massive body. A general-relativistic description of this system was given in Ref.~\cite{Barkov:2002mb}, while a quantitative description of the chaotic motion was given in Ref.~\cite{Barkov:2005}, for a particular regime in which the shells have a large hierarchy of mass scales. 
More recently, long-term evolutions of multiple shells have been presented in Ref.~\cite{Gaspar:2011zza}. We should stress that none of those works considered BH formation and critical phenomena in confining spaces. Confinement is crucial in our setup to force the shells to collide repeatedly, thus allowing small effects to build up in time.

In this work we extend the analysis of~\cite{Cardoso:2016wcr} to the AdS case, therefore removing the---now unnecessary---artificial reflecting surface that provided confinement. As expected, our results are in full qualitative agreement with Ref.~\cite{Cardoso:2016wcr}. In addition, we observe that this system of multiple shells in a confining ambient displays strikingly chaotic behavior.
We highlight the fact that the dynamics of such systems require only solving two {\it decoupled ODEs}.

%%%%%%%%%%%%%%%%%%%%%%%%%%%%%%%%%%%%%%%%%%%%%%%%%%
\section{Double-shell system\label{sec:Setup}}
%%%%%%%%%%%%%%%%%%%%%%%%%%%%%%%%%%%%%%%%%%%%%%%%%%

\begin{figure}[t]
\begin{center}
\includegraphics[width=0.3\textwidth]{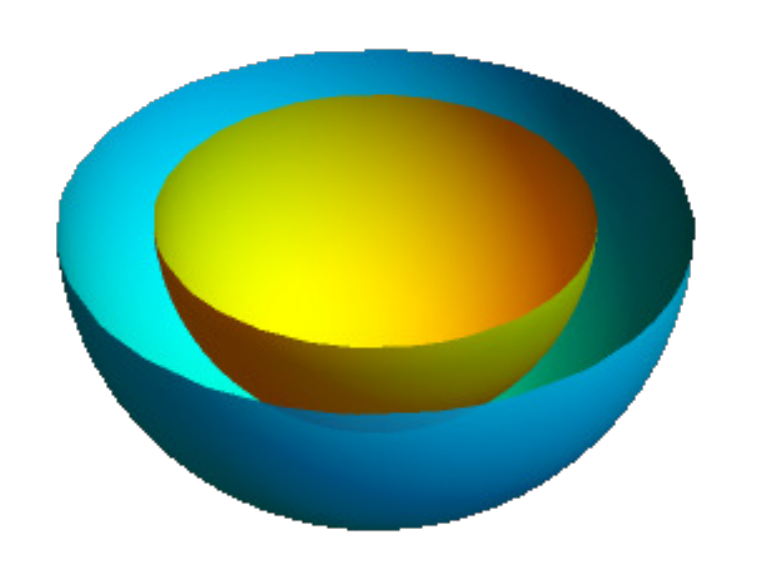}
\end{center}
\caption{Illustration of our setup: two concentric, spherically symmetric thin shells in an asymptotically AdS spacetime, shown above as two hemispherical domes for clarity.}
\label{fig:3Dshells}
\end{figure}

We consider the evolution of a spherically symmetric, asymptotically AdS spacetime with two concentric thin shells interacting only gravitationally (see Fig.~\ref{fig:3Dshells}). An exact (i.e., nonperturbative) description of the spacetime is obtained by gluing three Schwarzschild-AdS geometries along two timelike hypersurfaces.

%

%%%%%%%%%%%%%%%%%%%%%%%%%%%%%%%%%%%%%%%%%
\subsection{Evolving shells individually}
%%%%%%%%%%%%%%%%%%%%%%%%%%%%%%%%%%%%%%%%%

The time evolution of each shell can be followed individually up to the point that the two shells collide\footnote{At such events we have to make a choice for the subsequent evolution and this will be discussed below.}. The interior and exterior spacetimes are determined by Birkhoff's theorem to be described by AdS-Schwarzschild geometries,
\bea
ds^2&=&-f(r)dt^2+f(r)^{-1}dr^2+r^2d\Omega^2\,,\label{eq:metric}\\
f(r)&=&\left(1-\frac{2M}{r}+\frac{r^2}{l^2}\right)\,.
\eea
Here, $l$ is the (constant) AdS curvature. Once this is fixed, the only input needed is the gravitational mass of the interior and exterior regions. The induced metric on a shell of radius $r=R(\tau)$ is then
\be
d\sigma^2=-d\tau^2+R(\tau)^2d\Omega^2\,,
\label{eq:induced}
\ee
where $\tau$ denotes the shell's proper time and $d\Omega^2$ is the line element on the unit two-sphere.
We denote derivatives with respect to $\tau$ by an overdot.

The nonvanishing components of the extrinsic curvature are straightforwardly computed,
\beq
K_{\tau\tau}^\pm&=&-\frac{\dot{\beta}_\pm}{\dot{R}}\,, \qquad
K_{\theta\theta}^\pm=R\beta_\pm=K_{\phi\phi}^\pm/\sin^2\theta\,,\\
\beta_\pm&\equiv&\sqrt{\dot{R}^2+f_\pm(R)}\,,
\eeq
where $\pm$ applies to exterior and interior quantities, respectively.

Applying the Israel-Darmois junction conditions~\cite{Darmois:1927,Israel:1966rt}, a discontinuity of the extrinsic curvature signals the presence of a nonvanishing stress-energy tensor on the hypersurface given by
\be
S_{ij}=-(8\pi G)^{-1}\left([K_{ij}]-g_{ij}[K]\right)\,,
\label{eq:junction2}
\ee
where $[X]\equiv X_{+}-X_{-}$ denotes the jump of any quantity $X$ across the shell's surface and $K_{\pm}=\dot{\beta}_+/\dot{R}+2\beta_\pm/R$ is the trace of the extrinsic curvature.

We take the matter on the shell to be described by a perfect fluid,
\be
S_{ij}=(\rho+P)u_iu_j+Pg_{ij}\,,
\label{eq:perfectfluid}
\ee
where $u_i=\delta_i^\tau$ represents the fluid's $3$-velocity, $\rho$ its energy density, and $P$ its pressure.
By equating~\eqref{eq:junction2} to~\eqref{eq:perfectfluid} we thus find
\be
\rho=-\frac{1}{4\pi G R}[\beta]\,, \qquad
P=\frac{1}{8\pi G}\left(\frac{d[\beta]}{dR}+\frac{[\beta]}{R}\right)\,.
\label{eq:rho&P}
\ee

To close the system one must provide an equation of state relating the fluid's energy density and pressure.
We adopt, for simplicity, a linear equation of state $P=w\rho$, with $w$ a constant.
Consequently, integration of~\eqref{eq:rho&P} yields
\be
\rho=\frac{m\, l^{2w}}{4\pi G R^{2+2w}}\,,
\ee
with $m$ a constant, corresponding to the shell's invariant mass and $G$ denoting Newton's constant.
The inclusion of the factor $l^{2w}$ is a matter of convenience, preserving the mass dimension of $m$ for any choice of equation-of-state parameter $w$.

Inserting the above solution in Eq.~\eqref{eq:rho&P} one can obtain---after some massaging---a neat expression for the exterior gravitational mass, which for the pressureless case ($w=0$) reduces to a sum of the interior gravitational mass, the shell's kinetic energy and the shell's binding energy,
\be
M_+=M_-+\frac{m\, l^{2w}}{R^{2w}}\sqrt{\dot{R}^2+1+\frac{R^2}{l^2}-\frac{2M_-}{R}}-\frac{m^2l^{4w}}{2R^{1+4w}}\,.\label{eq:pre_motion}
\ee

However, for the purpose of studying the time evolution of the shell's radius it is more convenient to invert Eq.~\eqref{eq:pre_motion}, thus finding
\be
\dot{R}^2+V=0\,,\label{eq:motion}
\ee
where the radial effective potential is
\be
V=1+\frac{R^2}{l^2}-\frac{M_++M_-}{R}-\frac{(M_+-M_-)^2}{m^2}\left(\frac{R}{l}\right)^{4w}-\frac{m^2l^{4w}}{4R^{2+4w}}\,.\label{eq:potential}
\ee

As long as the energy density is positive and the equation-of-state parameter is in the range $-1/3\leq w\leq1$, all the standard energy conditions~\cite{Hawking:1973uf} are obeyed, namely, the null ($\rho+P\geq 0$), the weak ($\rho\geq 0, \rho+P\geq 0$), the strong ($\rho+P\geq 0, \rho+3P\geq 0$), and the 
dominant ($\rho\geq P\geq-\rho$) energy conditions.

For the two-shell system that we are interested in, we can use Eq.~\eqref{eq:motion} to follow the radius $R_{1,2}$ of the outermost and innermost shells. For the innermost shell we set $M_-=0$ and $M_+=M_2$, while the outermost will be described by $M_-=M_2$ and $M_+=M_1$. Because the proper time for the two shells will not coincide, in general, it is convenient to follow the evolution with respect to the Schwarzschild time coordinate $t$ for the region between the two shells. By considering Eqs.~\eqref{eq:metric} and~\eqref{eq:induced}, the Schwarzschild time $t$ is directly related to the proper time $\tau_{1,2}$ of the shell at radius $R_{1,2}$,
\be
\frac{dt}{d\tau_{1,2}}=\frac{\sqrt{f(R_{1,2})+\left(\dot{R}_{1,2}\right)^2}}{f(R_{1,2})}\,.
\ee
From this it immediately follows that the Schwarzschild time evolution of each shell is governed by
\be\label{time_evolution}
\left(\frac{dR}{dt}\right)^2= -\widehat{V}\equiv-\frac{f(R)^2 V}{f(R)-V}\,.
\ee

The system of ODEs~\eqref{time_evolution} is simple enough to be integrated using {\scshape Mathematica}'s built-in routine {\scshape NDSolve}. To integrate the equations we use the default settings of the routine, namely, a typical accuracy and a precision goal of 1 part in $10^8$. We checked that larger precisions do not change our results.

%%%%%%%%%%%%%%%%%%%%%%%%%%%%%%%%%%%%%%%%%%%%%%%%%%%
\subsection{Boundary conditions and shell crossing}
%%%%%%%%%%%%%%%%%%%%%%%%%%%%%%%%%%%%%%%%%%%%%%%%%%%
%
\begin{figure*}[ht]
%\begin{center}
\begin{tabular}{cc}
\includegraphics[width=0.45\textwidth]{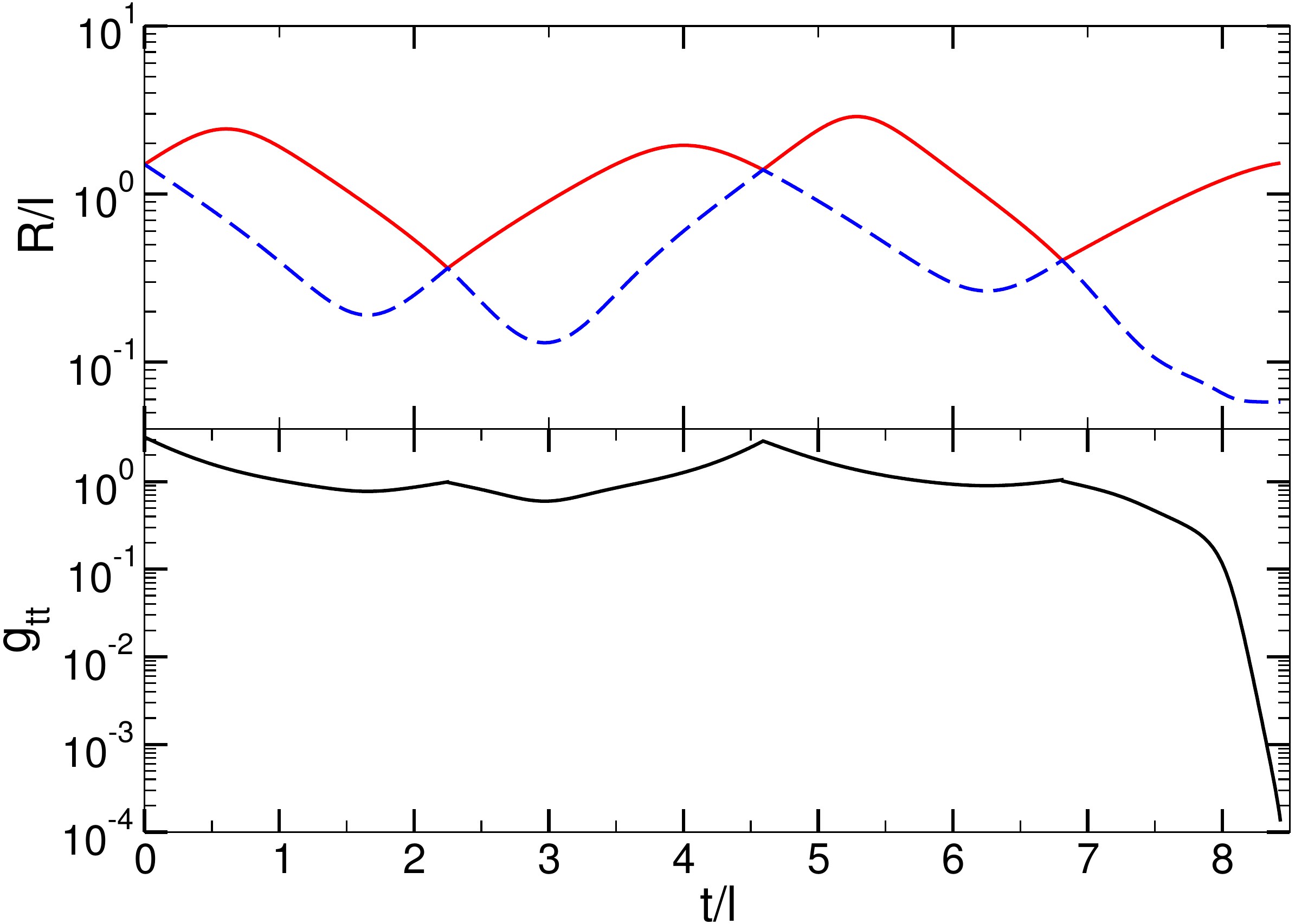}&
\includegraphics[width=0.45\textwidth]{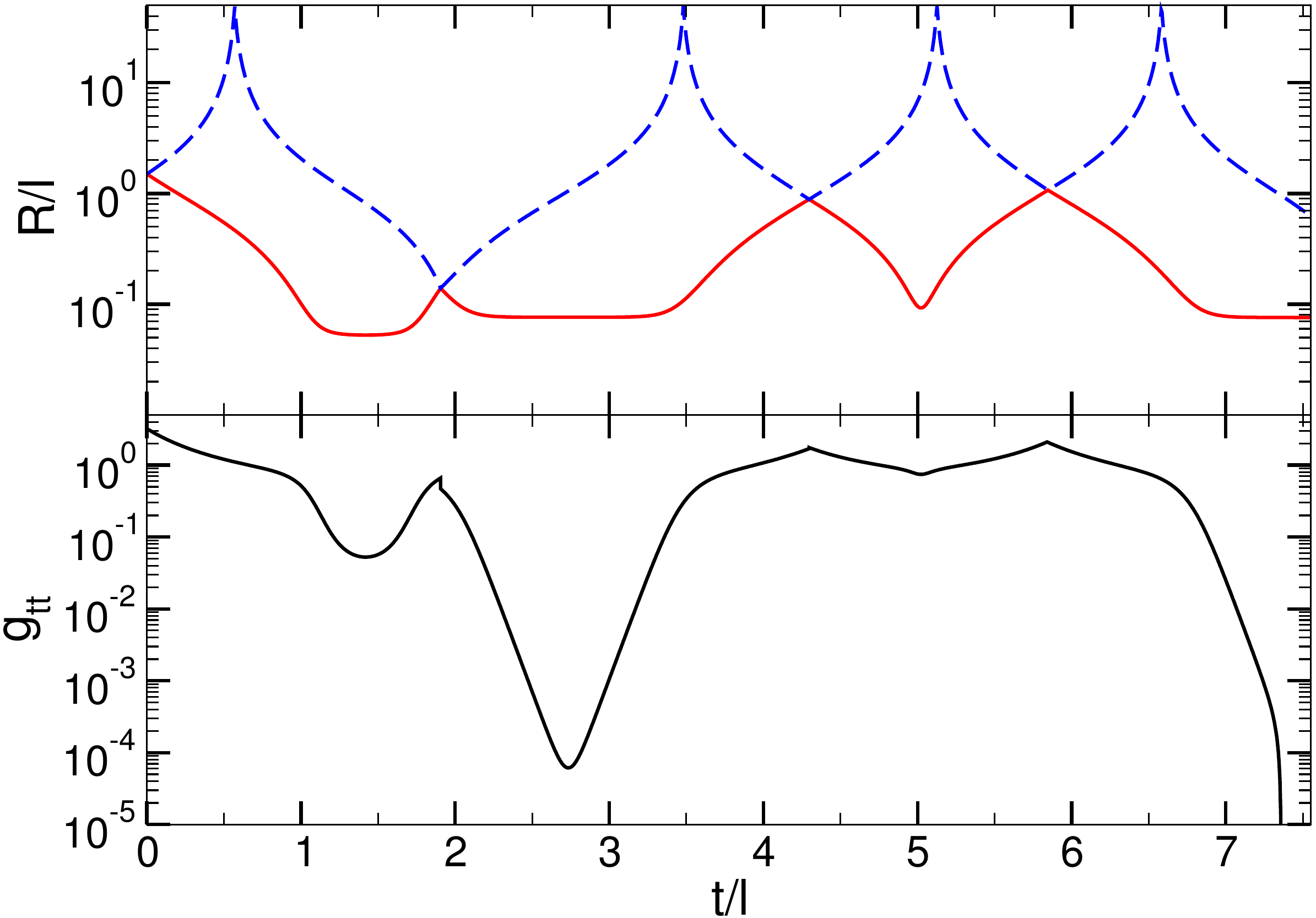}
\end{tabular}
%\end{center}
\caption{Evolution of the two-shell system in AdS for %$l=20,\,M_1=1, M_2=0.5,\,m_1=m_2=0.9$ and an initial shell radius $R_i=30$.
$M_1/l=0.05, M_2/l=0.025,\,m_1/l=m_2/l=0.0136$ and an initial shell radius $R_i/l=1.5$.
{\bf Left panels:} We consider $w_1=w_2=0.2$. The upper panel shows the position of the shells as a function of the Schwarzschild-like coordinate time $t$ measured between the shells. The solid (red) curve describes the motion of the outermost shell and the dashed (blue) curve the innermost. The lower panel shows the value of $g_{tt}=1-2M_{2}/R_{2}+R_2^2/l^2$ computed at the surface of the innermost shell. For this case, the innermost shell collapses after three crossings (meaning that it is the initially outermost shell that actually collapses). 
{\bf Right panels:} Same as the left but for $w_1=w_2=1$ and $m_1/l=m_2/l=0.0001125$, for which the outermost shell can reach $R\to \infty$ in a finite time.
\label{fig:Rvst}}
\end{figure*}
Due to the timelike asymptotic boundary of AdS, the shells can reach the boundary $R\to\infty$ within a finite time as measured by a static observer in the bulk. Due to the effective potential~\eqref{eq:potential} this can happen whenever $w>1/2$. To have a well-defined problem, we then impose perfectly reflecting boundary conditions ($\dot{R}\to -\dot {R}$) when $R\to\infty$. 
On the other hand, for $-1/3\leq w<1/2$~\footnote{For the case $w=1/2$, the sign of the potential at $R\to\infty$ depends on the sign of the combination $m^2-(M_+-M_-)^2$.} the effective potential~\eqref{eq:potential} has a maximum turning point after which the potential becomes positive, forbidding classical motion up to $R\to\infty$.\footnote{For some choices of the parameters, it was shown that for $w\leq 1/2$ even a single shell can oscillate between two finite radii~\cite{Mas:2015dra}. Oscillating single shell solutions can also be shown to exist even for $w>1/2$ if perfectly reflecting boundary conditions ($\dot{R}\to -\dot {R}$) are imposed at $R\to\infty$.}
Thus, as long as perfectly reflecting boundary conditions are imposed at the AdS boundary, the system will behave as a confined system for any value of $w$.

Following Ref.~\cite{Cardoso:2016wcr}, when the shells collide we consider them to be ``transparent,'' by keeping their 4-velocity continuous at the collision point and their invariant mass unchanged. However, there will still be an exchange of gravitational energy between the shells, such that the gravitational mass $M_2$, exterior to the innermost shell (and interior to the outermost shell), will change after each crossing. The gravitational mass $M_2$ after each collision has previously been computed from conservation of energy and momentum and can be found in Ref.~\cite{Ida:1999} [see their Eq.(3.18)].

%%%%%%%%%%%%%%%%%%%%%%%%%%%%%%%%%%%%
\section{Results\label{sec:Results}}
%%%%%%%%%%%%%%%%%%%%%%%%%%%%%%%%%%%%

%%%%%%%%%%%%%%%%%%%%%%%%%%%%%%%%%%%%
\subsection{Initial conditions}
%%%%%%%%%%%%%%%%%%%%%%%%%%%%%%%%%%%%

The dynamics of the two-shell system in AdS shows a very rich structure akin to what was found in Ref.~\cite{Cardoso:2016wcr} for shells inside a spherical box. Depending on the parameters and initial conditions, the system can collapse promptly, bounce at the AdS boundary or at some finite radius, and collapse after some crossings between the two shells, or it can oscillate forever. The outcome of the evolution depends, in general, sensitively on the parameters and initial conditions. 
As noted in the previous section, for example, only shells with $w>1/2$ will reach the AdS boundary at $R\to \infty$. However, the qualitative behavior of our results does not depend on the equation-of-state parameter $w$ as long as perfectly reflecting boundary conditions are imposed at $R\to \infty$. This is illustrated in Fig.~\ref{fig:Rvst}, where we compare the dynamics of double shells with $w$ smaller and larger than $1/2$. In the figure, the shell's radius is shown as a function of the coordinate time measured by a static observer located between the shells. In this example the shells collapse after three crossings.
For the case with $w>1/2$ the outermost shell reaches $R\to \infty$ in a finite time and bounces back, while for $w<1/2$ the shell bounces back at a finite radius.

We thus fix $w$ in our study. For concreteness, and to compare with the results of Ref.~\cite{Cardoso:2016wcr} we focus on shells that initially start at the same location $R(t=0)=R_i$, with one shell expanding and the other contracting. We also consider similar sets of initial conditions, namely,

%\noindent {\bf A.} $M_1=1, M_2=0.5,\,m_1=m_2=0.9,\,w_1=w_2=0.2$, keeping free the initial location of the shells, $R_i$;
\noindent {\bf (i)} $M_1/l=0.05, M_2/l=0.025,\,m_1/l=m_2/l=0.9\times20^{-1-2w}=0.0136,\,w_1=w_2=0.2$, keeping free the initial location of the shells, $R_i$;

%\noindent {\bf B.} $M_1=\delta, M_2=0.5\delta,\,m_1=m_2=0.9\delta,\,w_1=w_2=0.2,\,R_i=30$, keeping free the parameter $\delta$ that quantifies the energy content in the spacetime.
\noindent {\bf (ii)} $M_1/l=\delta, M_2/l=0.5\delta,\,m_1/l=m_2/l=0.9\times20^{-2w}\delta=0.2715\delta,\,w_1=w_2=0.2,\,R_i/l=1.5$, keeping free the parameter $\delta$ that quantifies the energy content in the spacetime.

Formation of a horizon can be signaled by the function $1-2M_{1,\,2}/R_{1,\,2}+R_{1,2}^2/l^2$ approaching zero or by checking that, when the shells are contracting, the shell's radius is smaller than the innermost turning point of the effective potential~\eqref{eq:potential}, since in this case the shell is not able to bounce back and avoid collapse.

%%%%%%%%%%%%%%%%%%%%%%%%%%%%%%%%%%%%%%%%%%%%%%%%%%%%%
\subsection{Delayed collapse and critical behavior}
%%%%%%%%%%%%%%%%%%%%%%%%%%%%%%%%%%%%%%%%%%%%%%%%%%%%%
%
\begin{figure}[ht]
\begin{center}
\includegraphics[width=0.45\textwidth]{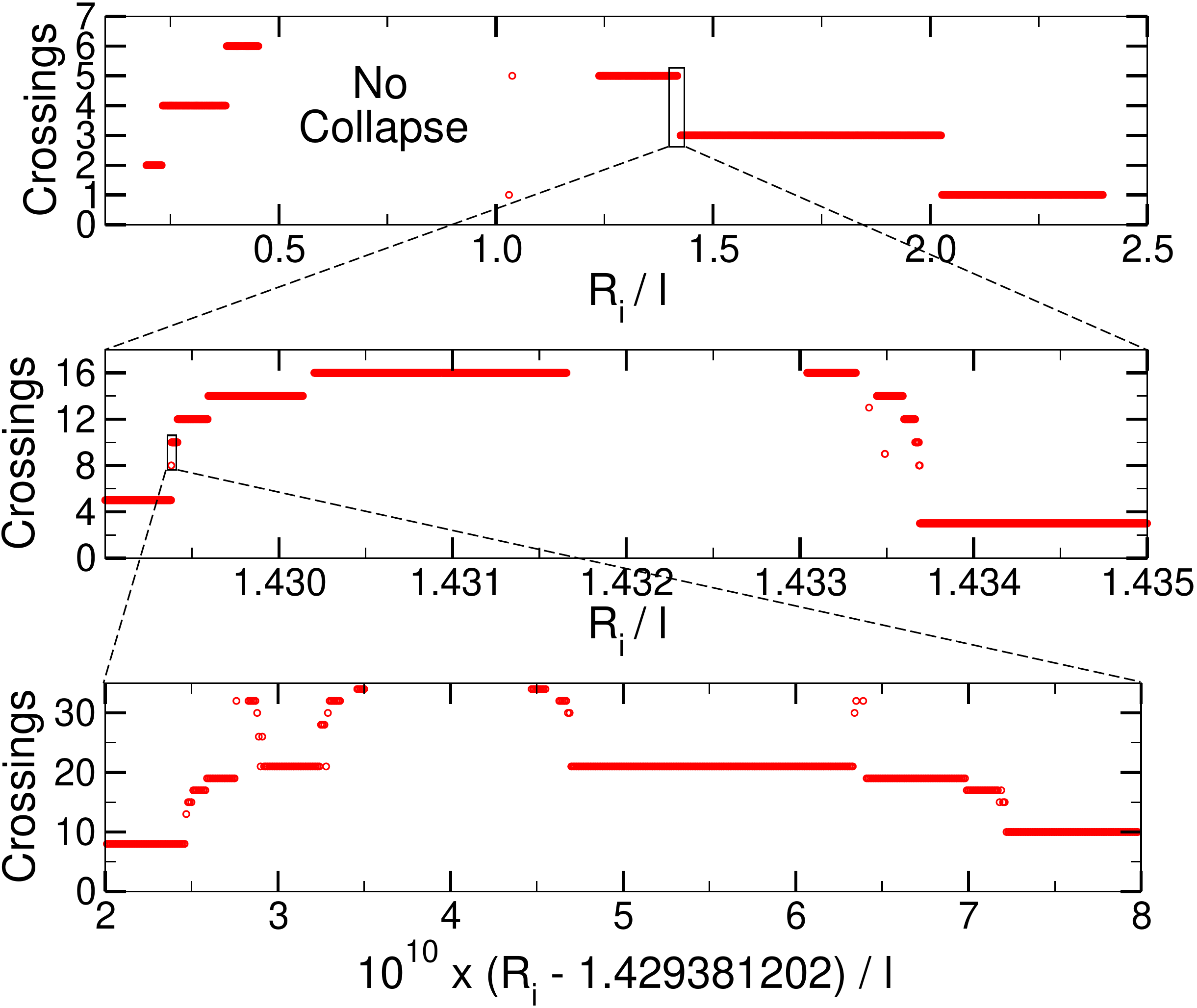}
\end{center}
\caption{Number of crossings between the two shells before collapse as a function of $R_i/l$, for type (i) initial conditions. We find regions where no collapse occurs, and around each critical point (i.e., when the number of crossings change) we observe a fractal-like structure where an arbitrarily large number of crossings occur, as shown in the center and bottom panels (cf. Fig. 3 in~\cite{Cardoso:2016wcr}).}
\label{fig:jvsRi_w02}
\end{figure}
The number of times the shells cross before collapsing depends sensitively on the parameters and the initial conditions. In Fig.~\ref{fig:jvsRi_w02} we show how the number of crossings before collapse changes when using type (i) initial conditions, i.e., varying the initial location $R_i$ of the shells. Between each transition there is a fractal-like structure: zooming-in between two plateaus, one finds that the structure resembles the top panel of Fig.~\ref{fig:jvsRi_w02} itself, as shown in the center and bottom panels. We also find a large region of $R_i$ for which there is no collapse.\footnote{Or, more precisely, for which there is no collapse up to $t\sim 10^3l$.} Odd-odd or even-even transitions are associated with critical points~\cite{Cardoso:2016wcr}, while parity transitions correspond to a mass gap in the mass of the formed BH. This behavior is completely analogous to what was found for the double-shell system inside a spherical box~\cite{Cardoso:2016wcr} (cf. Fig. 3 in~\cite{Cardoso:2016wcr}) and is similar to features found in the collapse of scalar fields in AdS.
Our results, together with the ones reported in~\cite{Cardoso:2016wcr}, confirm that collapse and critical behavior are triggered by the energy exchange between the two shells. The effect of the AdS boundary is to provide a natural confinement mechanism, allowing for the shells to cross multiple times.

\begin{figure}[ht]
\begin{center}
\includegraphics[width=0.45\textwidth]{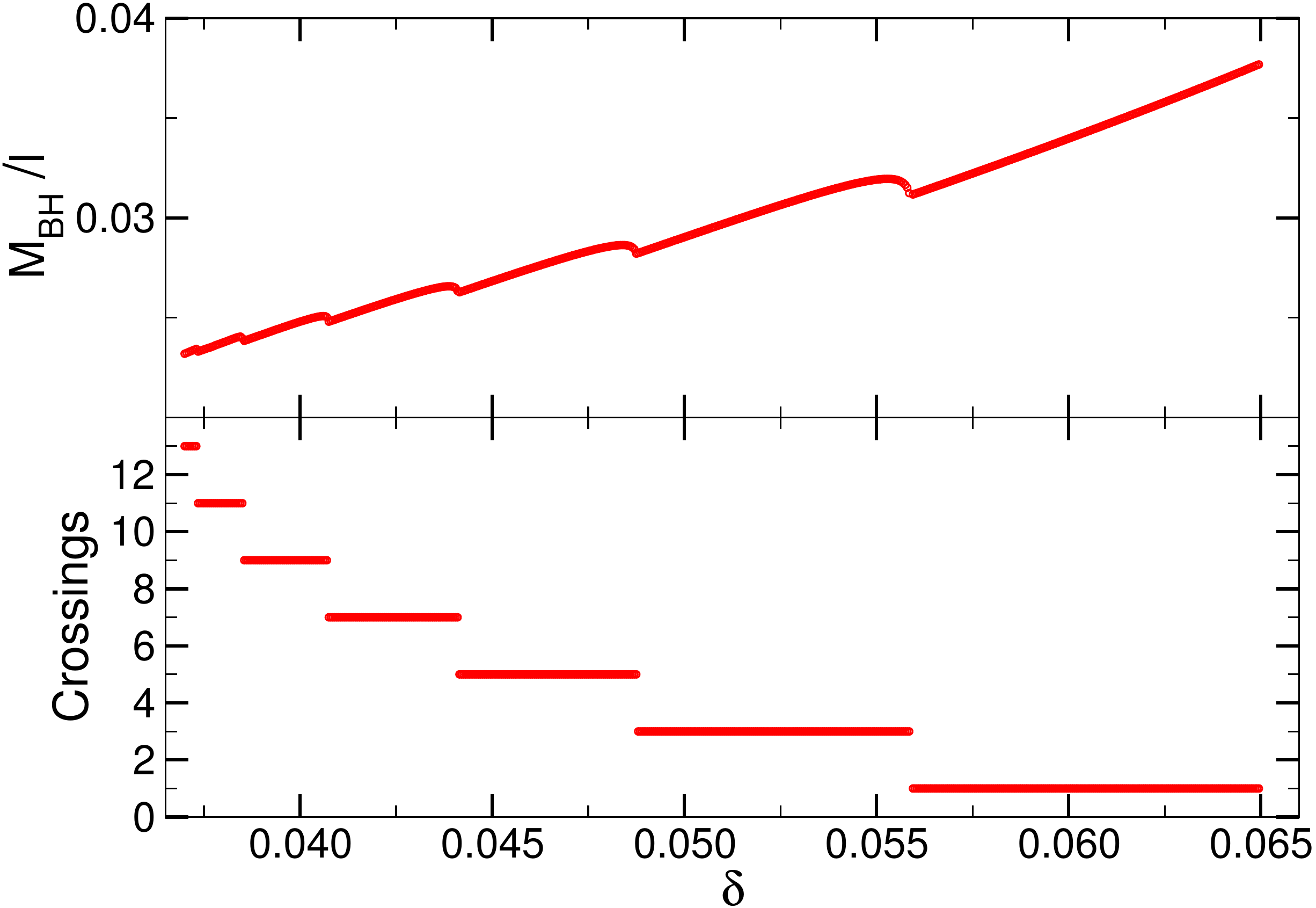}
\end{center}
\caption{The upper panel shows the BH mass as a function of the initial energy content $\delta$ for %$M_1/l=0.05\delta, M_2/l=0.025\delta,\,m_1/l=m_2/l=0.0136\delta,\,w_1=w_2=0.2,\,R_i/l=1.5$.
type (ii) initial data. In the lower panel we show the number of shell crossings, before each peak in the BH mass, which corresponds, from right to left, to $1,3,5,7,9,11,13$. Between each transition we find that, by fine-tuning $\delta$, an arbitrarily large number of crossings are, in principle, possible (see Fig. 3 in~\cite{Cardoso:2016wcr}). For $\delta\lesssim 0.0370$ we find no collapse (cf. Fig. 4 in~\cite{Cardoso:2016wcr}).}
\label{fig:Mvsdelta_w02}
\end{figure}
The critical behavior between odd-odd transitions (although not shown here, even-even transitions behave similarly) can be seen in Fig.~\ref{fig:Mvsdelta_w02}, where we plot the BH mass when a BH is formed as a function of $\delta$ for type (ii) initial data. The BH mass shows a typical critical behavior: it is a continuous function of $\delta$ to the left of the critical point, while its left derivative blows up at this point.\footnote{Although we do find a power-law scaling at the critical point, for the case shown in Fig.~\ref{fig:Mvsdelta_w02}, and within the accuracy of our results, we cannot actually rule out the possibility that the left derivative is finite at the critical point.} In the left neighborhood of these critical points, the BH mass is characterized by 
\be
M_{BH}-M_0\propto |\delta-\delta_*|^{\gamma}\,,
\ee
where $M_0$ is the BH mass at the critical point and $\gamma$ is a critical exponent.
%\footnote{The critical exponent $\gamma$ is universal in the sense that it does not depend on the physical parameter being varied to approach criticality. For example, for type A initial conditions we obtain $M_{BH}-M_0\propto |R_i-R_i^*|^{\gamma}$ with a similar value for $\gamma$.}. 
We find $\gamma\simeq 0.95\pm 0.05$ for the $1\to 3$ transition. The value obtained for the $3\to 5$ transition is consistent with this, but we do not have sufficient accuracy in all the branches to claim that the critical exponent is independent of the branch considered. We expect that the value of $\gamma$ depends on the type of matter, which in our case is encoded in the parameter $w$~\cite{Koike:1994sa}. For example, for $w=1$ we obtain a slightly smaller value $\gamma\sim 0.9$. It would be interesting to see if there are values of $w$ for which the exponent is closer to the one found in Ref.~\cite{Olivan:2015fmy}.

The critical exponent we obtain in this double-shell-AdS system is significantly larger than what was found for the case of two shells in a reflecting cavity. Namely, Ref.~\cite{Cardoso:2016wcr} found $\gamma\sim0.2$ for the exponent of the corresponding branch. This fact suggests that the cosmological constant plays an important role in the determination of the critical exponent for solutions that form a horizon with finite size. This result is consistent with the findings of Ref.~\cite{Cai:2016yxd}, which noted a similar suppression of the critical exponent in the collapse of massless scalar fields when the cosmological constant is turned off.

%%%%%%%%%%%%%%%%%%%%%%%%%%%%%%%%%%%%
\subsection{Chaotic behavior}
%%%%%%%%%%%%%%%%%%%%%%%%%%%%%%%%%%%%

So far, we have confirmed that the overall picture suggested by the study of two shells confined to a reflecting box in an otherwise flat space~\cite{Cardoso:2016wcr} still holds when one removes the mirror and places the system in AdS spacetime. One quantitative difference is the location of the outer turning point, which in the former case is fixed and in the latter is energy dependent. Now we analyze the chaotic nature of this system. This is manifest not only in the sensitivity of the number of crossings before collapse close to the critical points (see Figs.~\ref{fig:jvsRi_w02} and~\ref{fig:Mvsdelta_w02}) but also in the evolution of noncollapsing configurations.

\begin{figure*}[t]
\begin{center}
\begin{tabular}{cc}
\includegraphics[width=0.45\textwidth]{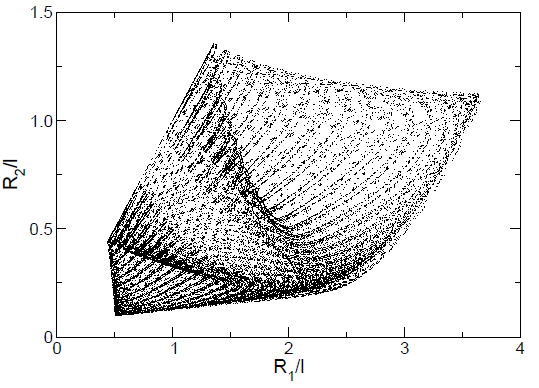}&
\includegraphics[width=0.45\textwidth]{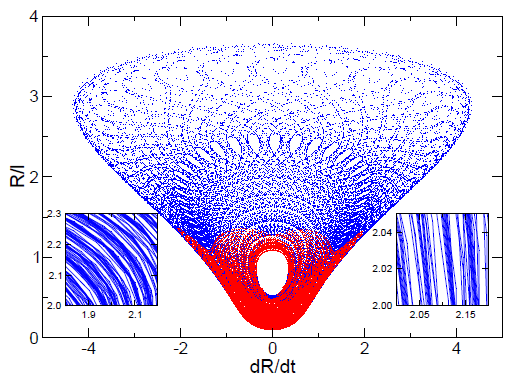}
\end{tabular}
\end{center}
\caption{Phase space of a noncollapsing solution for $M_1/l=0.03, M_2/l=0.015,\,m_1/l=m_2/l=0.00815,\,w_1=w_2=0.2,\,R_i/l=1$. {\bf Left:} Phase space in the $(R_1,R_2)$ plane with $R_1$ the radius of the outermost shell and $R_2$ the innermost shell's radius. The region covered by the orbits is thus obviously restricted to lie below the straight line $R_2=R_1$. {\bf Right:} Phase space in the $(R,dR/dt)$ plane. In red we show the innermost shell's orbits and in the blue the outermost. The inset plots show a zoom-in of the phase space.}
\label{fig:chaos}
\end{figure*}
\begin{figure}[ht]
\begin{center}
\includegraphics[width=0.45\textwidth]{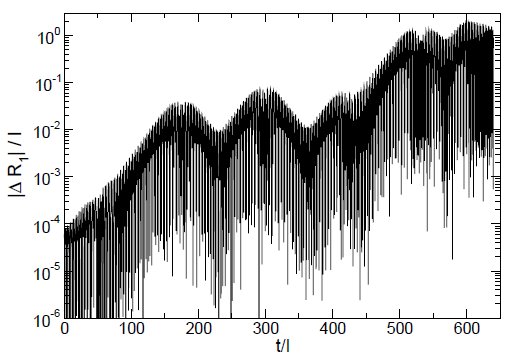}
\end{center}
\caption{Difference between the orbits of the outermost shell's radius for the parameters of Fig.~\ref{fig:chaos}, but with two initial conditions for $R_i$ that initially differ by $\Delta R_i/l=5\times 10^{-5}$. There is some evidence that the difference grows exponentially, indicating the presence of chaotic behavior.}
\label{fig:chaos_deltaR}
\end{figure}

Our results also show that there are ``islands of stability'' which are associated with oscillatory solutions. Performing a scan of these solutions, we find strong indications that some of these solutions display a mildly chaotic behavior. Plotting the orbits in the phase space of stable solutions we find that some of them show typical characteristics of chaotic systems. This is shown in Fig.~\ref{fig:chaos} where we plot the orbits of the double-shell system in the $(R_1,R_2)$ plane and $(R,dR/dt)$ plane for one of these solutions. It is apparent that a large area in phase space is covered by these orbits.

These configurations also show high sensitivity to initial conditions, typical of a chaotic system. This is illustrated in Fig.~\ref{fig:chaos_deltaR}. For two initial conditions for $R_i$ that initially differ by $\Delta R_i/l=5\times 10^{-5}$, the orbits in the phase space diverge exponentially from each other but always remain inside the same ``attractor.'' Since the rate of divergence is small ($|\Delta R_1/l| \sim e^{\lambda t/l}$ with a Lyapunov exponent $\lambda \simeq 0.04$ for the case shown in Fig.~\ref{fig:chaos_deltaR}), the chaotic nature is rather mild, and no distinction from a quasiperiodic motion is evident at early times.\footnote{Although quasiperiodic and weak chaotic motion can be difficult to distinguish in phase space, the fact that we find evidence for a positive Lyapunov exponent for some of these solutions is a clear signal of chaos.}

%In this respect, it is interesting to confront the Lyapunov exponent $\lambda$ we have obtained, with the conjectured upper bound by Maldacena, Shenker and Stanford~\cite{Maldacena:2015waa}: $\lambda\leq2\pi l T$, where $T$ is the temperature of the holographic dual quantum system. Taking this temperature to be the Hawking temperature of a Schwarzschild-AdS black hole with the same ADM mass $M_1$ as our (non-collapsing) shell system, one concludes that the conjectured bound is satisfied --- at least for this particular configuration. In fact, the Lyapunov exponent we obtained is far from saturating the holographic bound (by a factor of $\sim 200$).

\begin{figure}[ht]
\begin{center}
\includegraphics[width=0.45\textwidth]{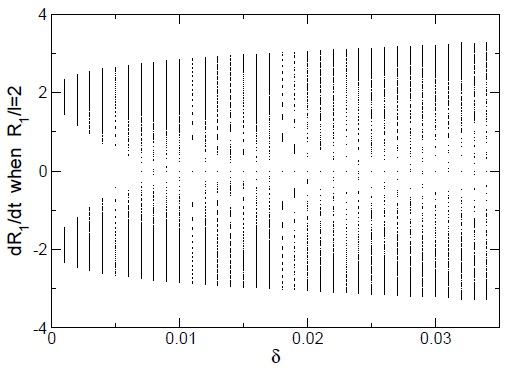}
\end{center}
\caption{Bifurcation diagram, illustrating the transition from (quasi)periodic to chaotic behavior of the double-shell system. We consider noncollapsing solutions obtained using initial data of type (ii) and plot the points where the orbit of a solution for a given $\delta$ intersects the line $R_1/l=2$ in the $(R_1,dR_1/dt)$-plane. For very small $\delta$ the solutions are nearly periodic. Increasing $\delta$ the system becomes ``increasingly'' chaotic, which is translated in a random and almost continuous distribution of points along a line of constant $\delta$.}
\label{fig:chaos_bifurcation}
\end{figure}

Another indication of chaotic behavior can also be found by constructing a bifurcation diagram when continuously varying some parameter of the system.
To do so we consider initial conditions of type (ii) and scan the parameter space of noncollapsing solutions, i.e., for $\delta\lesssim 0.0370$. In the limit $\delta\to 0$, the solutions are nearly periodic, but while increasing $\delta$, we find evidence that there is a transition to a chaotic motion. This is shown in Fig.~\ref{fig:chaos_bifurcation} where we consider the line $R_1/l=2$ in the $(R_1,dR_1/dt)$ plane and plot the points where the orbit of a solution for a given $\delta$ intersects this line. The transition from (quasi)periodic to nonperiodic motion when continuously varying a parameter of the system is a typical feature of chaotic systems. Evidence for a similar behavior was in fact also found for scalar fields in AdS~\cite{deOliveira:2012ac,Farahi:2014lta} (cf. Fig. 2 in Ref.~\cite{deOliveira:2012ac} and Fig. 3 in Ref.~\cite{Farahi:2014lta}).

%Depending on the parameters and initial conditions, different patterns can emerge, but we observe that chaotic behavior can be found quite generically for solutions that do not collapse.

%

We thus suggest the following picture. Our double-shell setup configures a chaotic system, at least for certain regimes of the parameter space. This typically occurs for noncollapsing configurations. The fractal-like structure observed around critical points with same-parity transitions is also likely because of chaotic behavior. As we have just demonstrated, two nearby initial configurations---call them ${\cal D}_{i,1}$ and ${\cal D}_{i,2}$---can diverge exponentially during the evolution. For certain choices of ${\cal D}_{i,1}$ we will see collapse after just a few orbits, but at that point the system starting from initial data ${\cal D}_{i,2}$ may be quite separated in phase space and thus continue evolving for a very long time before collapsing. This would explain the presence of the isolated points mentioned in Fig.~\ref{fig:jvsRi_w02}.

%%%%%%%%%%%%%%%%%%%%%%%%%%%%%%%%%%%%
\section{Conclusion\label{sec:Conc}}
%%%%%%%%%%%%%%%%%%%%%%%%%%%%%%%%%%%%

We have presented a study of a clean, physically appealing system composed of two concentric spherical thin shells in AdS spacetime, confirming that the overall picture inferred from the recent investigations of double shells in a reflecting box~\cite{Cardoso:2016wcr} faithfully represents this asymptotically AdS setting. Moreover, we uncovered interesting dynamics akin to chaotic systems and performed a first exploration of its main characteristics.

While the double-shell system may present some similarities with critical collapse of scalar fields in AdS (for instance, the existence of critical points and noncollapsing configurations), there certainly are marked differences. In the latter case, it is the nonlinearities, which are strongly felt when the field accumulates around the origin, that transfer energy to higher frequency modes and eventually lead to BH formation. On the other hand, shells (of finite mass) cannot shrink to arbitrarily small radius without collapsing into a BH. In multiple shell systems the transfer of energy instead comes from the shell-crossing events, which for ``transparent'' shells always flows from the outgoing shell to the ingoing shell~\cite{Nakao:1999}.

It would be interesting to determine whether some of the nonperiodic stable solutions for a scalar field in AdS, found in Ref.~\cite{Buchel:2013uba} display any kind of chaotic behavior, similar to the one found in this paper.
It would also be desirable to extract the Lyapunov exponent from an analytic treatment to compare with our numerical results. This would certainly add to a deeper understanding of the physics at play. However, it is not immediately clear how this can be achieved because one would have to analytically follow the evolution of the system over a time scale many times larger than the AdS light-crossing time, which must then incorporate many shell crossings.

The combination of simplicity and richness afforded by this setup provides an ideal test bed for explorations of gravitational collapse in confining geometries and its holographic dual interpretations. Further investigation is needed to determine if the physics at play in other problems of interest, e.g., collapse of scalar fields in AdS, is correctly captured by this elementary setting. At any rate, we hope the present study will shed new light on such problems, simultaneously serving as a powerful aid to a deeper understanding.

%%%%%%%%%%%%%%%%%%%%%%%%%%%%%
\begin{acknowledgments}
We would like to thank Maxim Barkov, Ben Craps, \'Oscar Dias, Roberto Emparan, Oleg Evnin, Bartomeu Fiol, Gavin Hartnett, Javier Mas and Juan Pedraza for useful comments and discussions.
R.B. acknowledges financial support from the FCT-IDPASC program through Grant No. SFRH/BD/52047/2012.
V.C. acknowledges financial support provided under the European Union's H2020 ERC Consolidator Grant ``Matter and strong-field gravity: New frontiers in Einstein's theory'' Grant Agreement No. MaGRaTh--646597. 
V.C. also acknowledges financial support from FCT under Sabbatical Fellowship No. SFRH/BSAB/105955/2014.
Research at Perimeter Institute is supported by the Government of Canada through Industry Canada and by the Province of Ontario through the Ministry of Economic Development $\&$
Innovation.
This work was supported by the H2020-MSCA-RISE-2015 Grant No. StronGrHEP-690904.
J.V.R. acknowledges financial support from the European Union's Horizon 2020 research and innovation programme under the Marie Sk\l{}odowska-Curie Grant Agreement No. REGMat-656882.
Funding for this work was partially provided by the Spanish MINECO under Project No. FPA2013-46570-C2-2-P.
\end{acknowledgments}
%%%%%%%%%%%%%%%%%%%%%%%%%%%%%

%\vskip 5mm

%%%%%%%%%%%%%%%%%%%%%%%%%%%%%
%\bibliographystyle{h-physrev4}

%%%%%%%%%%%%%%%%%%%%%%%%%%%%%

\end{document}